\documentclass[prl,twocolumn
]{revtex4}
\usepackage{graphicx}

\begin{document}

\title{The Temperature Echo of the Superconducting Phase Transition}

\author{V.~N.~Naumov$^1$}\email{vn@che.nsk.su}
\author{G.~I.~Frolova$^1$}
\author{T.~Atake$^2$}
\author{V.~V.~Nogteva$^1$}
\author{N.~A.~Nemov$^1$}
\author{M.~A.~Bespyatov$^1$}
\author{V.~G.~Potemkin$^1$}

\affiliation{$^1$Institute of Inorganic Chemistry, Siberian Branch of Russian Academy of Science, Novosibirsk, 630090, Russia \\
$^2$Materials and Structures Laboratory, Tokyo Institute of Technology, 4259 Nagatsuta-cho, Midori-ku, Yokohama, 226-8503, Japan}

\begin{abstract}
An analysis of experimental heat capacity at $T > T_c$ is presented for series of samples (R)Ba$_2$Cu$_3$O$_{6+x}$ (with $x$ close to optimal). For all samples the anomaly was discovered which occurred steadily in the interval 250--290~K (anomaly $T_{h}$). The anomaly $T_{h}$ looks like a phase transition anomaly. It was shown that the anomaly $T_{h}$ correlates with superconducting anomaly $T_c$, temperatures $T_{h}$ and $T_{c}$
being connected by the ratio $T_{h}\approx 3T_c$. The anomaly $T_{h}$ is interpreted as the origination of pairing the charge carriers. Anomalies at $T\approx 3T_c$ were also detected in heat capacity of low temperature superconductors Hg and Nb$_3$Ge.
\end{abstract}

\date{\today}%
\maketitle

\section{Introduction}
At present the study of peculiarities in different properties of cuprate superconductors (HTSC's) in a normal state attracts a special attention. It results from a tendency for understanding a pseudogap phenomenon revealing itself as a lowering the electron
density of states well above the superconducting temperature $T_{c}$. The problem is discussed whether or not both the pseudogap and the superconductivity are the phenomena of the same origin. It is believed that the solution is of fundamental importance for the clarification of high temperature superconductivity mechanism in cuprate systems. This explains a great number of works in the field (see 
references in reviews~\cite{1}, \cite{2}, \cite{3}, \cite{4}, \cite{5}, \cite{6}, \cite{7}, \cite{8}).

Now it is not clear if the pseudogap line is only crossover, Ref.~\cite{9}, \cite{10},
or it is accompanied by the hidden change of order parameter symmetry, Ref.~\cite{11}, \cite{12}, \cite{13}, \cite{14}. It seems to be actual to investigate the heat capacity in order to clarify the question about the possible critical behavior of these objects in the temperature region of normal state.

In this work we present results of such investigation for series of 90 K samples (R)Ba$_2$Cu$_3$O$_{6+x}$ (where R means Y or 
Gd, Tm, Ho) with $x$ close to the optimal value. Making use of elaborated before technique for separating the heat capacity into different contributions, Ref.~\cite{15}, we examine experimental data in a wide temperature interval of normal state. An anomaly was discovered which occurs steadily in the range 250--290~K ($T_{h}$ anomaly). It was noted that as $T_c$ changes, $T_h$ also changes so that the relation $T_{h}\approx 3T_c$ takes place~\cite{16}, \cite{17}. Now we present additional evidences of such interrelation.

\section{Experimental}
The experimental heat capacity $C_p(T)$ above $T_{c}$ can be
presented by the following expression~\cite{18}:
\[
C_p(T)=C_{harm}(T)+\gamma T +A(T-T_0)^\alpha+\delta C(T).
\]
The term $C_{harm}(T)$ describes a harmonic lattice part. The term $\gamma T$ describes a linear electron part and a linear anharmonic contribution. The term $A(T-T_0)^\alpha$ describes a possible nonlinear in $T$ anharmonic contribution and a possible contribution of low temperature wing of anomaly resulted from ordering oxygen~\cite{18}. The first three terms describe regular contributions to the heat capacity. The term $\delta C(T)$ describes the possible localized anomalies. Subtracting the regular contributions from the experimental heat capacity we obtain the component $\delta C(T)$. Regular contributions are smooth and monotone which can introduce no localized in temperature peculiarities when subtracting. They were determined from the experimental heat capacity as described in~\cite{15}, \cite{18}.

\begin{figure}[tbh]
\includegraphics[width=0.45\textwidth]{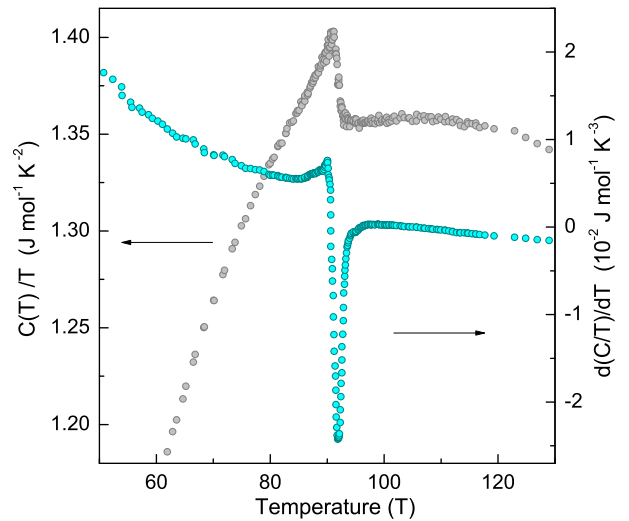}
\caption{\label{fig:fig1} The functions $C(T)/T$ and $d(C/T)/dT$ for the compound YBa$_2$Cu$_3$O$_{6.95\pm{0.02}}$ on the interval of superconducting phase transition.}
\end{figure}

\begin{figure}[tbh]
\includegraphics[width=0.45\textwidth]{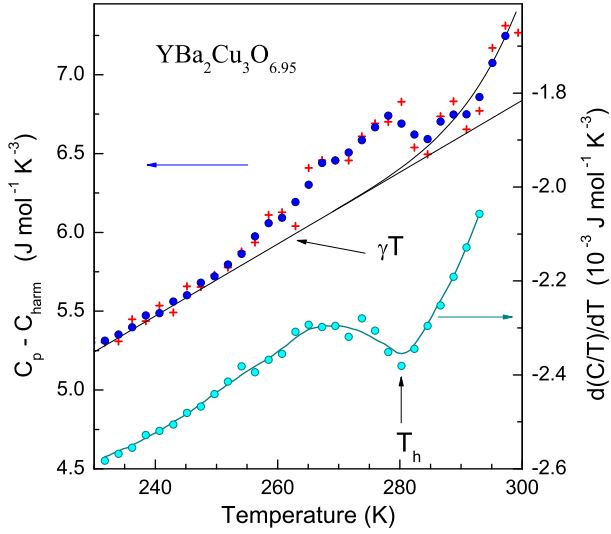} 
\caption{\label{fig:fig2} The functions $C_p - C_{harm}$ and $d(C/T)/dT$ ) for the compound YBa$_2$Cu$_3$O$_{6.95\pm{0.02}}$ in the range 230--290 K.}
\end{figure}

As an example, the data for YBa$_2$Cu$_3$O$_{6.95}$ can be considered. The superconducting phase transition of this sample takes place at $T_c \approx 92$~K (Fig.~\ref{fig:fig1}).
An anomaly was observed at $\sim 277$ K (anomaly $T_{h}$, Fig.~\ref{fig:fig2}). The temperatures $T_{h}$ and $T_{c}$ are related by the ratio $T_h/T_c\approx 3$. The anomaly $T_{h}$ looks like a phase transition anomaly.

\begin{figure}[tbh]
\includegraphics[width=0.45\textwidth]{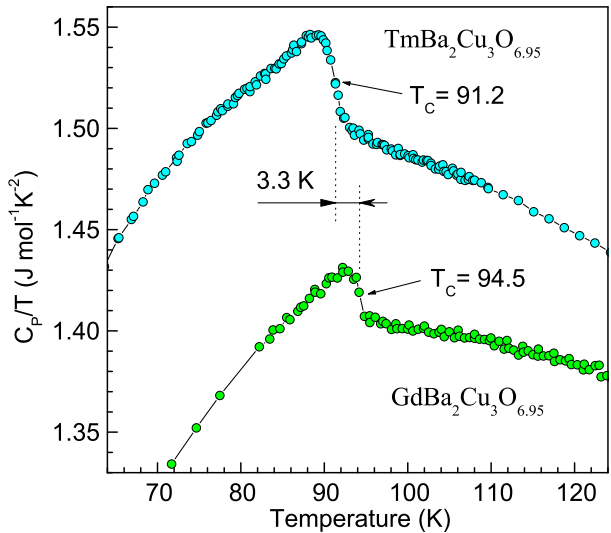} 
\caption{\label{fig:fig3} The function $C_p/T$ on the interval of
superconducting phase transition for compounds TmBa$_2$Cu$_3$O$_{6.95\pm{0.02}}$ and GdBa$_2$Cu$_3$O$_{6.95\pm{0.02}}$.}
\end{figure}

Another example which gives evidence to such a relationship
between $T_c$ and $T_h$ is presented below for two samples
TmBa$_2$Cu$_3$O$_{6.95}$ and GdBa$_2$Cu$_3$O$_{6.95}$. These samples have
been synthesized by identical way, and measurements of their heat
capacity have been carried out in the same calorimeter with the
same operating conditions. The temperatures of superconducting
transitions differ distinctly from each other: $T_c$ = 91.2 K for
TmBa$_2$Cu$_3$O$_{6.95}$ and $T_c = 94.5$ K for GdBa$_2$Cu$_3$O$_{6.95}$ (Fig.~\ref{fig:fig3}).

\begin{figure}[tbh]
\includegraphics[width=0.45\textwidth]{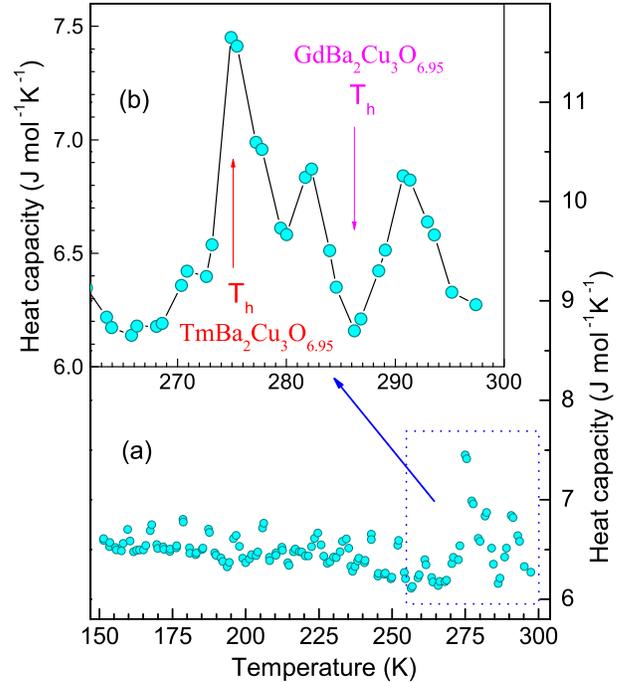} 
\caption{\label{fig:fig4} The difference of molar heat capacities of compounds
TmBa$_2$Cu$_3$O$_{6.95\pm{0.02}}$ and GdBa$_2$Cu$_3$O$_{6.95\pm{0.02}}$: a) on the interval 150--300 K, b) the same in a large scale, 262--300~K.}
\end{figure}

If indeed the temperatures $T_h$ and $T_c$ are related by the
ratio $T_h$/$T_c\approx 3$, then a peak in heat capacity of
TmBa$_2$Cu$_3$O$_{6.95}$ can be expected at $T_h \approx 274$~K and a
peak in heat capacity GdBa$_2$Cu$_3$O$_{6.95}$ can be expected at
$T_h \approx 284$~K. To check this prediction the difference
$C_{dif}$ of their molar heat capacities has been derived in the
range 150--300 K (Fig.~\ref{fig:fig4}).

In the obtained difference, all the errors sources of which are in
the experimental technique and sources of which are caused by a
presence of possible off-controllable impurities are practically
excluded. As a background, only the smooth contribution from a
Shottky anomaly in heat capacity of TmBa$_2$Cu$_3$O$_{6.95}$ and the
characteristic spread remain in this difference. It is seen in
Fig.~\ref{fig:fig4} that indeed, on this background, there is a maximum at
$\sim 274$~K corresponding to the expected peak $T_h$ for
TmBa$_2$Cu$_3$O$_{6.95}$ and there is a minimum at $\sim 285$~K,
corresponding to the expected peak $T_h$ for GdBa$_2$Cu$_3$O$_{6.95}$.
In Fig.~\ref{fig:fig4}b the difference $C_{dif}$ is presented in a large
scale.

From experiments with the heat capacity of two samples HoBa$_2$Cu$_3$O$_{6+x}$ we obtained one further example supported the presence of anomaly at $T \approx 3T_c$ (anomaly $T_h$). The sample HoBa$_2$Cu$_3$O$_{6.95}$ was prepared by the method of
powder-calcination. The high purity reactants Ba(CO$_3$),
Ho$_2$O$_3$ and CuO (99.99, 99.9 and 99.99 \% respectively)
were mixed in stoichiometric proportions and ground with an agate
mortar and a pestle. The mixture was two times heated at 1170 K
for 12 h. The sample was finally sintered in oxygen atmosphere at
1230 K for 24 h, and then cooled down to 620 K. After annealing at
620 K for 3 h in oxygen atmosphere, the sample was quenched
rapidly to room temperature. The product was identified by X-ray
powder diffraction which provided no evidence of extra phases
other than the phase of intended orthorhombic structure. The
oxygen content was determined by iodometric titration as
corresponding to the formula HoBa$_2$Cu$_3$O$_{6.95}$.

The heat capacity $C_p(T)$ has  been measured in the range 8--300~K. There was observed a characteristic of the sample: its superconducting anomaly was split into two peaks, with $T_c = 86$~K and $T_c = 90$~K. For extracting the anomalous component from the heat capacity $C_p(T)$, the regular background has been subtracted. As a regular
background, the heat capacity of another sample HoBa$_2$Cu$_3$O$_{6+x}$ with the oxygen index assumed to be about 6.5--6.6 was used. This sample was synthesized by the identical method and differed only by the annealing regime. The measurement of its heat capacity was carried out in the same calorimeter and at the same conditions within the same temperature range. No anomalies were observed over all temperature range.

\begin{figure}[tbh]
\includegraphics[width=0.46\textwidth]{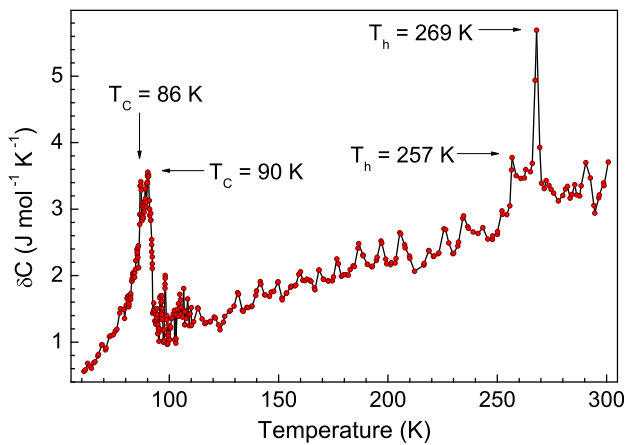}
\caption{\label{fig:fig5} Anomalies in heat capacity of HoBa$_2$Cu$_3$O$_{6.95}$ in the range 50--300 K.}
\end{figure}

It is obvious that the obtained difference $\delta C(T)$ between heat capacities of the above compounds  contains the possible anomalies in $C_p(T)$ of HoBa$_2$Cu$_3$O$_{6.95}$ and a small difference between their regular electron and phonon components. It is presented in Fig.~\ref{fig:fig5}.

\begin{figure}[tbh]
\includegraphics[width=0.42\textwidth]{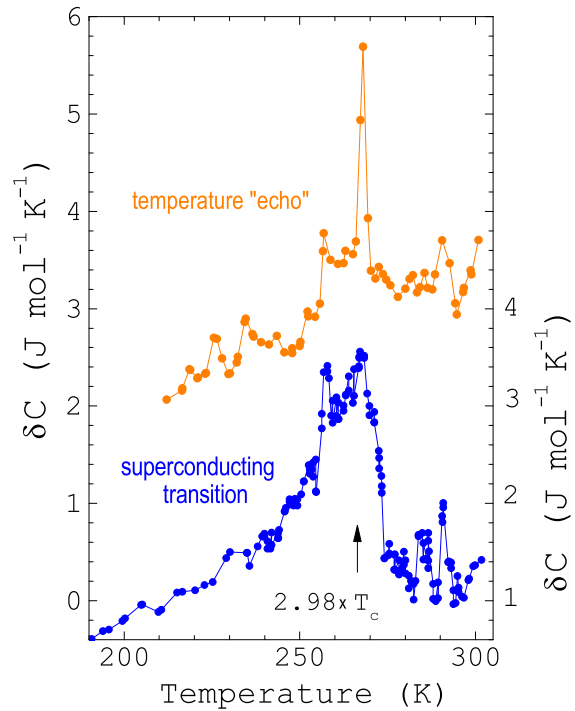} 
\caption{\label{fig:fig6} The difference of molar heat capacities of two compounds
HoBa$_2$Cu$_3$O$_{6+x}$ in the range 200--300~K. The lower curve is the superconducting anomaly shifted to higher temperatures by factor 3 for comparison of the shapes.}
\end{figure}

It should be noted that in addition to the split superconducting
anomaly, there is the two peak anomaly in the normal state region
(two peaks $T_h = 257$~K and $T_h = 269$~K), the relation
$T_h/T_c$ being equal 3 to a good accuracy for both peaks. It
looks as though the superconducting phase transition is imaged
into other temperature interval. In {Fig.~\ref{fig:fig6}} the difference $\delta C(T)$ is
presented on the interval 200-300 K in a large scale.

In given examples the localized anomaly $T_{h}$ was separated
using three different ways for subtracting the regular
contributions from the experimental heat capacity. Ordinary for 90~K compounds (R)Ba$_2$Cu$_3$O$_{6+x}$ with $x$ close to the optimal
value the anomaly $T_{h}$ is revealed in the range 250--290~K. As
$T_c$ changes, $T_h$ also changes so that the relation
$T_{h}\approx 3T_c$ remains the same.

As a rule, for different samples the amplitude of this anomaly
varies from 0.4 to 3.6 J mol$^{-1}$K$^{-1}$ which is 0.15--1.3~\% of
total heat capacity. In individual cases the amplitude of anomaly
$T_h$ is comparable to the amplitude of anomaly $T_c$. The spread
of experimental points in the interval 100--300~K varies from 0.02\% up to 0.1\% depending on an amount of a sample and on calorimeter and installation characteristics for concrete measurements.

Results of heat capacity investigations point out that the shape and the amplitude of superconducting anomaly $T_c$ depend substantially on details of synthesis of 90~K samples, Ref.~\cite{19}, \cite{20}. We believe that the shape and the amplitude of anomaly $T_h$ also depend substantially on details of synthesis. Therewith it is possible that they differ even for the samples with the same $x$.

\section{Discussion}
The anomaly $T_{h}$ in heat capacity looks like a phase transition anomaly. Its shape, in occasion, resembles the shape of superconducting anomaly $T_{c}$. The above examples show that the higher is $T_c$, the higher is $T_h$ and there exists the relationship between the values $T_h$ and $T_c$: $T_h \approx 3T_c$. The anomaly $T_c$ is as though mapped onto other temperature interval in form of anomaly $T_{h}$. This correlation between temperatures $T_h$ and $T_c$ points to a relationship between the $T_h$ process and the superconductivity. We have called such correlation the "temperature echo".

From the above correlation between the anomalies $T_h$ and $T_c$
it might be concluded that $T_h$ process like the $T_c$
process is connected with the changes in electron subsystem and
reflects the onset of superconductivity. The change of electron
density of states $\delta N(E_F)$ resulted from $T_h$ - anomaly
can be estimated by relationship $\Delta \gamma = \Delta S/T_h$ 
($\delta N(E_F)$ is proportional to $\Delta \gamma$, here $\gamma$
is the coefficient of electron heat capacity, $\Delta S$ is the
entropy of anomaly $T_h$). The estimation for HoBa$_2$Cu$_3$O$_{6.95}$
exhibiting rather high anomaly $T_h$ among the series of our
investigated compounds shows that $\Delta\gamma$ is less than 5\%
of $\gamma$. Such change is difficult to detect directly which
might explain why there are no evidences of anomaly $T_h$ in
literature.

\begin{figure}[tbh]
\includegraphics[width=0.45\textwidth]{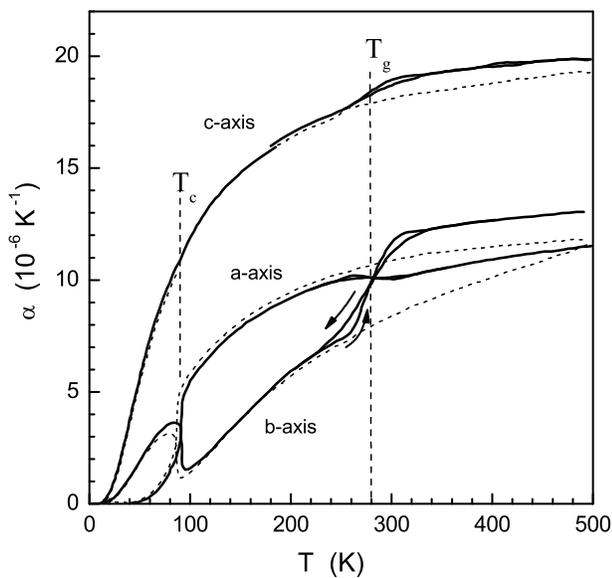} 
\caption{\label{fig:fig7} The coefficient of thermal expansion $\alpha (T)$ along the three
orthorhombic axes for the single crystal YBa$_2$Cu$_3$O$_{6.95}$ (Figure from article P.~Nagel et. al., Ref.~\cite{22}).}
\end{figure}

The occurrence of some process at $T \approx 3T_c$ is supported by the anomalies in some other properties of superconductors (R)Ba$_2$Cu$_3$O$_{6+x}$ (see for example Ref.~\cite{21}, \cite{22}). In the work~\cite{21} the step-like anomaly of ultrasound velocity in
superconductors GdBaSrCu$_3$O$_{7-x}$ ($T_c = 82$~K) was noted around temperature $T_g = 245$~K indicating a pronounced lattice hardening (note that $T_g \approx 3T_c$ !). In the work~\cite{22} the coefficient of thermal expansion $\alpha (T)$ along the three orthorhombic axes was measured for the single crystal YBa$_2$Cu$_3$O$_{6.95}$. Along the axes $a$ and $b$ the anomaly at $T_c$ was observed and also the anomaly at the temperature $T_g = 290$~K was observed, Fig.~\ref{fig:fig7}. Note that $T_g \approx 3T_c$ ! These results indicate that in a lattice subsystem the same relationship between the temperatures $T_g$ and $T_c$ occurs.

Taking into account the strong electron - phonon coupling in (R)Ba$_2$Cu$_3$O$_{6+x}$ superconductors (see Ref.~\cite{4}, \cite{5}, \cite{6}, \cite{7}) one can expect that at any change of electron characteristics, the phonon characteristics inevitably change. Thus it looks like the anomalies at $T_g$ in lattice subsystem and the anomaly at $T_h$ in electron subsystem are of the same nature.

The peculiarities of different properties observed in temperature interval above $T_c$ are ordinarily put down to the pseudogap phenomena, see reviews \cite{1}, \cite{2}, \cite{3}, \cite{5}, \cite{7}. This pseudogap $\Delta_d$ (doped) is the reduced density of states intrinsic to the energy band structure of HTSCs and revealing itself as peculiarities of some physical properties in vicinity of temperature $T^{*}$ in underdoped range. Now the corresponding line on the phase diagram $T^{*}(x)$ is well known~\cite{1}, \cite{2}, \cite{3}, \cite{5}, \cite{8}: the values $T^{*}$ are lowering when $x$ is growing. The results of numerous studies show that crossing this line results in something other than phase transition and is unrelated to superconductivity~\cite{5}, \cite{7}.

The coexistence of two phenomena in electron subsystem at $T_h$ and $T_c$ leads us to the following assumption. If $T_c$ is the temperature of origin of superconducting gap $\Delta_c$, then $T_h$ can be identified with the temperature of origin of another pseudogap $\Delta_p$ (pairing) and the $T_h$-process can be identified with the pairing of charge carriers which results in suppression of spectral weight in the quasiparticle spectrum. This process means the onset of superconductivity. On the phase diagram it will be presented by the line $T_h \approx 3T_c(x)$.

So, one can concede that there exist two different pseudogaps in the density of states of superconductors (R)Ba$_2$Cu$_3$O$_{6+x}$. One is the pseudogap $\Delta_p$ connected with the onset of superconductivity and revealed itself in heat capacity as a phase transition at temperature multiple of $T_c$. Another pseudogap, well known one, is the pseudogap $\Delta_d$, appeared in different properties of HTSCs in vicinity of line $T^{*}(x)$ as
something other than phase transition.

The discovered phenomenon (temperature echo) might represent the thermodynamic evidence of the origination of superconducting pseudogap $\Delta_p$ at temperature $T_h \approx 3T_c(x)$.

\section{Supplement to Discussion}
If the cuprate superconductors with the strong elect\-ron-phonon coupling exhibit the anomaly $T_h$ in heat capacity (temperature echo of superconducting phase transition) then one can expect that the low temperature superconductors with strong electron - phonon coupling also exhibit the similar $T_h$ anomaly in heat capacity.

\begin{figure}[tbh]
\includegraphics[width=0.40\textwidth]{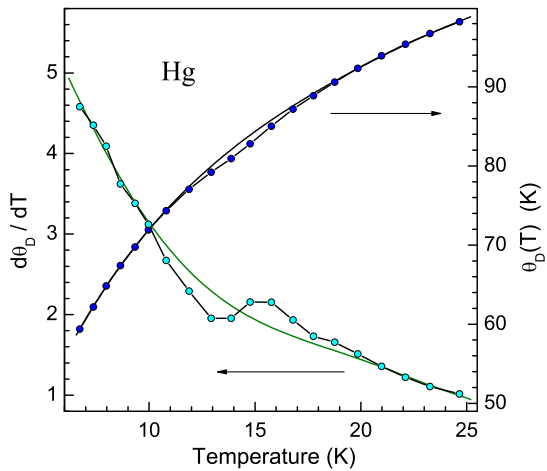} 
\caption{\label{fig:fig8} The Debye temperature $\Theta_D(T)$ of mercury (Hg) and the
derivative $d\Theta_D/dT$ (calculated from data Ref.~\cite{23}).}
\end{figure}

\begin{figure}[tbh]
\includegraphics[width=0.40\textwidth]{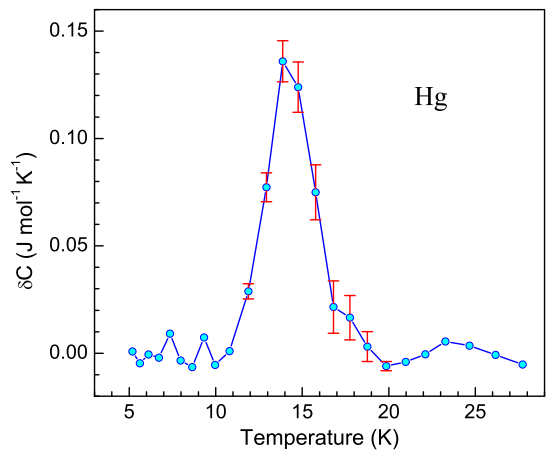} 
\caption{\label{fig:fig9} The anomaly in heat capacity of mercury (Hg) at temperature $\sim 3T_c$ (calculated from data Ref.~\cite{23}).}
\end{figure}

The high accuracy experimental data on heat capacity of mercury Hg ($T_c \approx 4.2$~K) have been examined~\cite{23}. In Fig.~\ref{fig:fig8} the Debye temperature $\Theta_D(T)$ and the derivative $d\Theta_D/dT$ are presented. Indeed, it occurs that there exists the anomaly at temperature $\sim 3T_c$. The peak value of anomalous heat capacity amounts up to 2.1\% of regular heat capacity, mean deviation of experimental points in the range 5--20~K being less then 0.01\% (Fig.~\ref{fig:fig9}).

\begin{figure}[!tbh]
\includegraphics[width=0.45\textwidth]{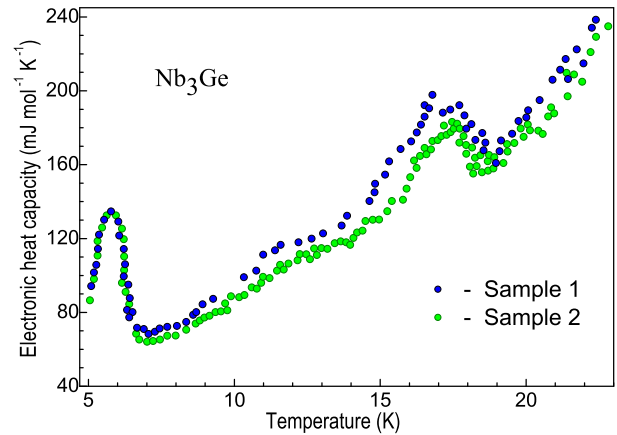} 
\caption{\label{fig:fig10} The sum ($\gamma T + \delta C$) of electronic heat capacity and
anomalous part of heat capacity of Nb$_3$Ge.}
\end{figure}

Another example is a compound of $A15$ structure Nb$_3$Ge. The samples were produced in the following way. Niobium (Nb) 99.999\% pure and germanium (Ge) 99.9999\% pure were ground to powder and mixed in the wanted proportion 75\% of Nb and 25\% of Ge, and tablets were pressed from this mix. A tablet was hung in induction furnace in helium atmosphere and fused. A drop of fusion fell on a cold base (fast quenching $\sim 10^4$~K/\,sec). The heat capacity for two samples (unannealed and annealed at 600--700$^\circ$C) was measured. The lattice heat capacity was approximated by the expression
\[
C_{latt}(T) = (A + BT) T^3.
\]
It was subtracted from the experimental heat capacity and the result of subtracting (the sum of electron heat capacity and anomalous part ($\gamma T + \delta C$)) is presented in
Fig.~\ref{fig:fig10}. The superconducting anomaly can be seen at $\sim 6$ K and another anomaly at temperature $\sim\!3T_c$ (at $\sim\! 18$ K) can be seen as well.

One can suppose that the anomaly $T_h$ in heat capacity of compounds (R)Ba$_2$Cu$_3$O$_{6.95}$ and the anomaly at $T \approx 3T_c$ in heat capacity of above low temperature superconductors are of the same nature. So the discovered phenomenon might be intrinsic to the superconductors with strong electron-phonon coupling, and the temperature echo might appear not only in cuprate superconductors.

\section{Summary}
The anomaly $T_h$ in heat capacity of superconductors (R)Ba$_2$Cu$_3$O$_{6+x}$ with $x$ close to optimal, was discovered at $T > T_c$ (250--290~K). The anomaly $T_h$ looks like a phase transition anomaly.

The correlation was found between the anomalies $T_h$ and $T_c$, $T_h \approx 3T_c$. This correlation was called the phenomenon of temperature "echo".

The $T_h$ process and the superconducting phase transition ($T_c$ process) are related to each other: they touch on both the electron and the phonon subsystem and can be attributed to superconductivity. The $T_h$ process might be interpreted as the origination of pairing the charge carriers. The corresponding lowering of electron density of states might be identified as the superconducting pseudogap $\Delta_p$.

Two low temperature superconductors Hg and Nb$_3$Ge were shown to exhibit the anomaly in heat capacity at $T \approx 3T_c$. It may be that analogous phenomenon occurs in other
superconductors with strong coupling.

\section{Acknowledgments}
The work is supported by Russian Foundation of Basic Research
(grant No: 03-03-32446).

\end{document}